\documentclass[superscriptaddress,twocolumn]{revtex4}
\usepackage{graphicx}
\usepackage{color}

\renewcommand{\vec}[1]{\mbox{\boldmath $#1$}}

\begin{document}

\title{Disappearance of nuclear deformation in hypernuclei: 
a perspective from a beyond-mean-field study} 

\author{H. Mei}
\affiliation{Department of Physics, Tohoku University, Sendai 980-8578, Japan}

\author{K. Hagino}
\affiliation{Department of Physics, Tohoku University, Sendai 980-8578, Japan}
\affiliation{Research Center for Electron Photon Science, Tohoku University, 
1-2-1 Mikamine, Sendai 982-0826, Japan}

\author{J.M. Yao}
\affiliation{
FRIB/NSCL Laboratory, Michigan State University, East Lansing, 
Michigan 48824, USA} 
\affiliation{School of Physical Science and Technology,
             Southwest University, Chongqing 400715, China}

\author{T. Motoba}
\affiliation{Laboratory of Physics, Osaka Electro-Communications University,
             Neyagawa 572-8530, Japan}
\affiliation{Yukawa Institute for Theoretical Physics, Kyoto University, Kyoto 606-8502, Japan }


\begin{abstract}
The previous mean-field calculation 
[Myaing Thi Win and K. Hagino, Phys. Rev. C{\bf 78}, 054311 (2008)] 
has shown that the oblate 
deformation in $^{28,30,32}$Si disappears when a $\Lambda$ particle 
is added to these nuclei. 
We here investigate this phenomenon by taking into account 
the effects beyond the mean-field approximation. 
To this end, we employ the microscopic particle-rotor model 
based on the covariant density functional theory. 
We show that the deformation of $^{30}$Si does not 
completely disappear, even though it 
is somewhat reduced, after a $\Lambda$ particle is added
if the beyond-mean-field effect is taken into account. 
We also discuss the impurity effect of $\Lambda$ particle 
on the electric quadrupole 
transition, and show that an addition of a $\Lambda$ 
particle leads to a reduction in the $B(E2)$ value, 
as a consequence of the reduction in the deformation parameter. 
\end{abstract}

\maketitle

\section{Introduction} 

The nuclear deformation is one of the most important concepts in 
nuclear physics \cite{BM75,RS80}. 
Whereas only those states with good angular momentum 
are realized in the laboratory, atomic nuclei can be deformed in the 
intrinsic frame, in which the rotational symmetry is spontaneously 
broken. This idea nicely explains the existence of rotational bands  
as well as enhanced electric transitions within the 
rotational bands in many nuclei. 
Theoretically, the nuclear deformation is intimately related to the mean-field 
approximation \cite{RS80,BHR03}, but there have also been recent 
attempts to describe the characteristics of deformed nuclei using 
symmetry preserved frameworks 
\cite{Dytrych13,Maris15,Stroberg16,Jansen16,GAB18}. 

In this paper, we shall 
discuss the nuclear deformation of single-$\Lambda$ 
hypernuclei 
\cite{Feshbach76,Zofka80,Zhou07,MH08,Schulze10,Lu11,Isaka11,Myaing11,Isaka12,Isaka13,Isaka14,Lu14,Cui15}, where a $\Lambda$ particle is added to atomic nuclei. 
See Refs. \cite{HT06,HY16,GHM16} for reviews on hypernuclei. 
A characteristic feature of hypernuclei is that a $\Lambda$ particle 
does not suffer from the Pauli principle of nucleons, and thus its 
wave function can have a large probability at 
the center of hypernuclei. 
This may significantly affect the structure of atomic nuclei. 

In the history of hypernuclear studies, 
when the experimental data of strangeness-exchange 
$(K^-,\pi^-)$ reactions came out from CERN, 
Feshbach proposed the concept of ``shape polarizabilty", 
that is, a possible change of 
nuclear radius and deformation induced by the hyperon participation \cite{Feshbach76}. 
Subsequently, \v{Z}ofka carried out Hartee-Fock calculations for hypernuclei 
to analyze such effects on even-even nuclei with $Z=N$ and $A<40$ \cite{Zofka80}.
He found that the relative change in quadrupole deformation should
be maximum at $^9_{\Lambda}$Be and $^{29}_{~\Lambda}$Si in the 
$p$-shell and $sd$-shell, respectively, although the expected change 
was not so large (only of the order of 1-4\% in $sd$-shell). See also Ref. \cite{Zhou07}. 
In modern light of nuclear structure studies, however, such 
response to the $\Lambda$ participation depends sensitively on
the nuclear own properties such as softness and potential shape. 
As a matter of fact, 
based on the relativistic mean-field (RMF) theory, 
it was argued that the nuclear deformation 
may disappear in some nuclei, such as $^{12}$C and $^{28,30,32}$Si, 
when a $\Lambda$ particle is added to these nuclei \cite{MH08}. 
That is, those deformed nuclei turn to be spherical hypernuclei 
after a $\Lambda$ particle is put in them. 
See also Refs. \cite{Lu11,Isaka11} for a similar conclusion. 
It has been shown that a softness of the potential energy surface 
in the deformation space is a primary cause of this 
phenomenon \cite{Schulze10}. 

In general, one expects a large fluctuation around the minimum 
when a potential surface is soft against deformation. 
This effect can actually 
be taken into account by going beyond the mean-field 
approximation with the generator coordinate method (GCM) \cite{RS80,BHR03}. 
In addition, one can also apply the angular momentum and the 
particle number projections to a mean-field wave function, in which 
these symmetries are spontaneously broken. 
Such calculations have been performed recently not only for ordinary 
nuclei \cite{BH08,Egido10,Yao10,Yao11,Yao14,Bally14,Egido16} 
but also for hypernuclei \cite{MHY16,Wu17,Cui17}. 
We shall here apply the beyond-mean-field calculations to 
a typical soft hypernucleus, as the most appropriate theoretical 
treatment for the dynamical shape fluctuation.

The aim of this paper is then  
to asses the effect beyond the mean-field approximation 
on the phenomenon of disappearance of nuclear deformation, which takes place 
in hypernuclei whose potential surface is soft. 
A similar work has been carried out 
with the anti-symmetrized molecular dynamics \cite{Isaka16}. 
Here, we instead employ the 
microscopic particle-rotor model based on the covariant density functional theory 
\cite{Mei14,Xue15,Mei15,Mei16,Mei17}, in which 
the $\Lambda$ particle motion is coupled to the core wave functions described 
with the beyond-mean-field method. 

The paper is organized as follows. In Sec. II, we briefly summarize the 
microscopic particle-rotor model. 
In Sec. III, we apply this framework to the $^{31}_{~\Lambda}$Si hypernucleus, 
for which the disappearance of deformation has been found in 
the mean-field approximation, and 
discuss the impurity effect of $\Lambda$ particle 
on the structure of the soft nucleus, $^{30}$Si. 
We then summarize the paper in Sec. IV. 

\section{Microsocpic Particle-Rotor Model}

We consider in this paper a single-$\Lambda$ hypernucleus. 
The Hamiltonian for this system reads, 
\begin{equation}
H=T_\Lambda+H_{\rm core}+\sum_{i=1}^{A_C}v_{N\Lambda}(\vec{r}_\Lambda,\vec{r}_i), 
\end{equation}
where 
$T_\Lambda$ is the kinetic energy of the $\Lambda$ particle and 
$H_{\rm core}$ is the many-body Hamiltonian for the core nucleus, 
whose mass number is $A_C$. 
$v_{N\Lambda}(\vec{r}_\Lambda,\vec{r}_i)$ is the nucleon-$\Lambda$ ($N\Lambda$) 
interaction, in which 
$\vec{r}_\Lambda$ and $\vec{r}_i$ denote the coordinates of the $\Lambda$ 
particle and of the nucleons, respectively. 

In the microscopic particle-rotor model, the total wave function 
for the system is described as 
\begin{eqnarray}
\Psi_{JM_J}(\vec{r}_\Lambda,\{\vec{r}_i\})
&=&\sum_{j,l}\sum_{n,I}  {\cal R}_{jlnI}(r_\Lambda) \nonumber \\
&&\times
[{\cal Y}_{jl}(\hat{\vec{r}}_\Lambda)\otimes\Phi_{nI}(\{\vec{r}_i\})]^{(JM_J)},
\label{wf}
\end{eqnarray}
where $J$ is the angular momentum of the hypernucleus and $M_J$ is 
its $z$-component in the laboratory frame. 
${\cal R}_{jlnI}(r_\Lambda)$ and ${\cal Y}_{jlm_j}(\hat{\vec{r}}_\Lambda)$ 
are the radial and the spin-angular wave functions for 
the $\Lambda$ particle, with $j$, $m_j$, and $l$ being the 
total single-particle momentum and its $z$-component, and the orbital 
angular momentum, respectively. 
In Eq. (\ref{wf}), 
$\Phi_{nIM}(\{\vec{r}_i\})$ is a many-body wave function for the core 
nucleus, satisfying 
$H_{\rm core}|\Phi_{nIM}\rangle=\epsilon_{nI}|\Phi_{nIM}\rangle$, 
where $I$ and $M$ are the total angular momentum and its $z$-component 
in the laboratory frame 
for the core nucleus, and $n$ is the index to distinguish different states 
with the same $I$ and $M$. 

The radial wave function, 
${\cal R}_{jlnI}(r_\Lambda)$, in Eq. (\ref{wf}) is obtained by solving 
the coupled-channels equations given by, 
\begin{eqnarray}
0&=&
\langle[{\cal Y}_{jl}(\hat{\vec{r}}_\Lambda)\otimes\Phi_{nI}(\{\vec{r}_i\})]^{(JM_J)}
|H-E_J|\Psi_{JM_J}\rangle, \\
&=&
\left[T_\Lambda(jl)+\epsilon_{nI}-E_J\right]
{\cal R}_{jlnI}(r_\Lambda) \nonumber \\
&&~~~~~+\sum_{j',l'}\sum_{n',I'}V_{jlnI,j'l'n'I'}(r_\Lambda)\,{\cal R}_{j'l'n'I'}(r_\Lambda),
\label{cc}
\end{eqnarray}
with 
\begin{equation}
V_{jlnI,j'l'n'I'}(r_\Lambda)=
\left\langle jlnI\left|\sum_{i=1}^{A_C}v_{N\Lambda}(\vec{r}_\Lambda,\vec{r}_i)
\right|j'l'n'I'\right\rangle, 
\end{equation}
where 
$|jlnI\rangle\equiv 
\left|[{\cal Y}_{jl}(\hat{\vec{r}}_\Lambda)\otimes\Phi_{nI}(\{\vec{r}_i\})]^{(JM_J)}
\right\rangle$. 

In the microscopic particle-rotor model, 
the core wave functions, $\Phi_{nIM}$, are constructed 
with the generator coordinate method 
by superposing projected Slater determinants, $|\phi_{IM}(\beta)\rangle$, 
as, 
\begin{equation}
|\Phi_{nIM}\rangle = \int d\beta\,f_{nI}(\beta)|\phi_{IM}(\beta)\rangle, 
\label{proj_wf}
\end{equation}
where $\beta$ is the quadrupole deformation parameter 
and $f_{nI}(\beta)$ is the weight function. 
In writing this equation, for simplicity, we have assumed that 
the core nucleus has an axially symmetric shape. 
Here, 
$|\phi_{IM}(\beta)\rangle$ is constructed as 
\begin{equation}
|\phi_{IM}(\beta)\rangle = \hat{P}^I_{M0}\hat{P}^N\hat{P}^Z|\beta\rangle, 
\label{mf}
\end{equation}
where $|\beta\rangle$ is the wave function 
obtained with a constrained mean-field method 
at the deformation $\beta$, and 
$\hat{P}^I_{M0}$, $\hat{P}^N$, and $\hat{P}^Z$ are the 
operators for the angular momentum projection, the particle number 
projection for neutrons, and that for protons, respectively. Notice 
that the $K$-quantum number is zero in $\hat{P}^I_{M0}$ because of the 
axial symmetry of the wave function, $|\beta\rangle$. 
The weight function, $f_{nI}(\beta)$, in Eq. (\ref{proj_wf}) is determined 
with the variational principle, which leads to the Hill-Wheeler 
equation \cite{RS80}, 
\begin{eqnarray}
&&\int d\beta'\,\left[\langle\phi_{IM}(\beta)|H_{\rm core}|\phi_{IM}(\beta')\rangle 
\right. \nonumber \\
&&\left.~~~~~~~-\epsilon_{nI}\langle\phi_{IM}(\beta)|\phi_{IM}(\beta')\rangle 
\right]f_{nI}(\beta')=0.
\end{eqnarray}
Notice that by setting $f_{nI}(\beta)=\delta(\beta-\beta_0)$ in 
Eq. (\ref{proj_wf}), one can also obtain the projected energy 
surface, $E_{J}(\beta_0)$, after solving the 
coupled-channels equations, Eq. (\ref{cc}) \cite{Xue15}. 
(In this case, there is only one single state, $n=1$, in the 
core nucleus for each $I$.) 

See Refs. \cite{Mei14,Xue15,Mei15,Mei16,Mei17} for more details on 
the framework of the microscopic particle-rotor model. 

\section{Deformation of the $^{31}_{~\Lambda}$S\lowercase{i} hypernucleus} 

We now apply the microscopic particle-rotor model to 
$^{31}_{~\Lambda}$Si as a typical example of hypernuclei which show 
the disappearance of nuclear deformation in the mean-field approximation. 
To this end, we employ the relativistic point-coupling model. 
For the core nucleus, $^{30}$Si, we use the PC-F1 \cite{PC-F1} 
parameter set, while we use PCY-S4 \cite{PCY-S4} 
for the $N\Lambda$ interaction. 
As we have shown in Ref. \cite{Mei16}, 
the dependence of the results on the choice of 
the $N\Lambda$ interaction would not be large 
and the conclusion of the paper 
will remain the same, at least qualitatively, even if we use another 
set of the PCY-S interaction. 
The pairing correlation among the nucleons in the core nucleus is 
taken into account in the BCS approximation with a contact pairing 
interaction with a smooth energy cutoff, 
as described in Ref. \cite{PC-PK1}. 
We generate the reference states, $|\beta\rangle$, in Eq. (\ref{mf}) 
by expanding the single-particle wave functions on a harmonic oscillator 
basis with 10 major shells. The coupled-channels calculations are also 
solved by expanding the radial wave functions, 
${\cal R}_{jlnI}(r_\Lambda)$, on the spherical harmonic oscillator basis 
with 18 major shells. In the coupled-channels calculations, we include 
the core states up to $n_{\rm max}=2$ and $I_{\rm max}=6$. 

\begin{figure}[t]
\includegraphics[clip,width=9cm]{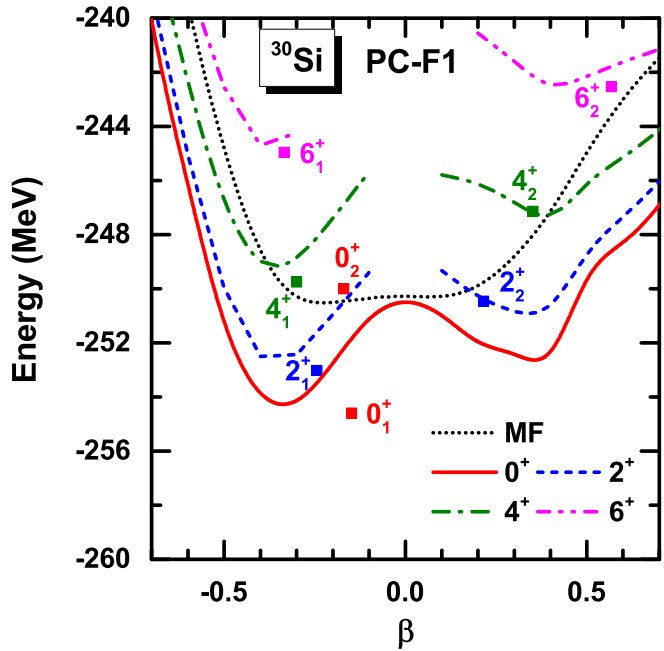}
\caption{
The projected energy curves for the $^{30}$Si nucleus as a function of the 
quadrupole deformation parameter, $\beta$. 
The mean-field energy curves are also shown by the dotted lines 
for a comparison. 
The filled squares indicate the energy of the GCM solutions, which 
are plotted at their average deformation. 
}
\end{figure}

\begin{figure}[t]
\includegraphics[clip,width=9cm]{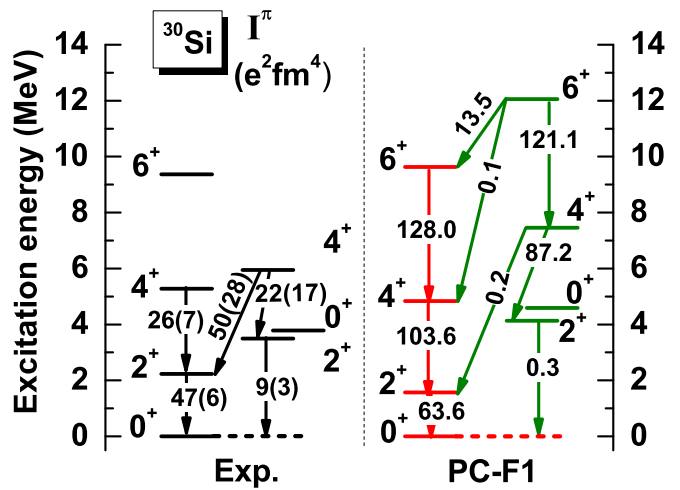}
\caption{
The low-lying spectrum of the $^{30}$Si nucleus obtained with 
the GCM method with the covariant density functional 
with the PC-F1 set. 
The arrows indicate the electric quadrupole (E2) transition 
strengths, plotted in units of $e^2$fm$^4$. 
These are compared with the experimental data taken from 
Ref. \cite{Basunia10}. 
}
\end{figure}

We first discuss the results for the core nucleus, $^{30}$Si. 
Figure 1 shows the potential energy curves for $^{30}$Si 
as a function of the deformation parameter, $\beta$. 
The energy curve in the mean-field approximation 
shows a shallow minimum at $\beta=-0.22$ (see the dotted line), 
which is similar to the 
energy curve for $^{28}$Si 
shown in Ref. \cite{MH08} obtained with the RMF 
theory with the meson-exchange 
NLSH parameter set \cite{NLSH}. 
For the projected energy curves, 
this calculation yields a well pronounced 
oblate minimum. For instance, 
for the $0^+$ configuration, the minimum appears at 
$\beta=-0.35$. 
The results of the GCM calculations 
for the spectrum as well as the $E2$ transition probabilities 
are shown in Fig. 2. 
The energy of each state is plotted also in Fig. 1, at the position of 
the mean deformation for each state. 
These calculations reproduce the experimental data reasonably well, 
even though 
the $B(E2)$ values for the intraband and the interband 
transitions are somewhat overestimated and underestimated, respectively. 

\begin{figure}[t]
\includegraphics[clip,width=8cm]{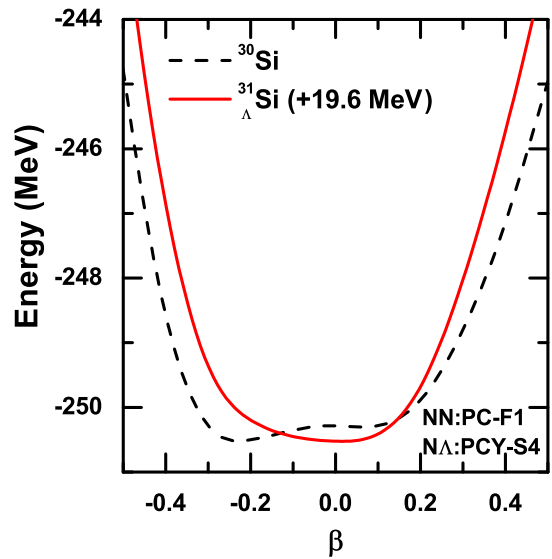}
\caption{
The potential energy curves in the mean-field approximation 
for the $^{30}$Si nucleus (the dotted line) and for the 
$^{31}_{~\Lambda}$Si hypernucleus (the solid lines). 
The energy curve for 
$^{31}_{~\Lambda}$Si is shifted in energy as indicated in the 
figure in order to compare 
with that for $^{30}$Si. 
}
\end{figure}

\begin{figure}[t]
\includegraphics[clip,width=8cm]{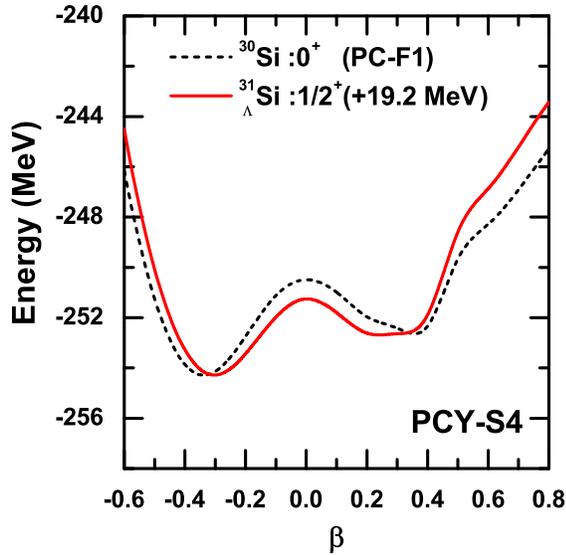}
\caption{
The projected energy curve for the $J^\pi=1/2^+$ configuration of the 
$^{31}_{~\Lambda}$Si hypernucleus (the solid lines). 
This is shifted in energy as indicated in the figure in order to compare 
with the energy curve for the core nucleus, $^{30}$Si (the dotted lines). 
}
\end{figure}

Let us now put a $\Lambda$ particle onto the $^{30}$Si nucleus and 
discuss the structure of the $^{31}_{~\Lambda}$Si hypernucleus. 
Fig. 3 shows the potential energy surface in the mean-field 
approximation, in which the curve for the hypernucleus (the solid line) is 
shifted in energy as indicated in the figure 
so that the energy of the absolute minima becomes the same as 
that for the core nucleus (the dotted line). 
One can see that the potential minimum is shifted 
from the oblate shape to the spherical 
shape by adding a $\Lambda$ particle to $^{30}$Si. 
As we have mentioned, the same phenomenon 
has been found also with another relativistic interaction, that is, the 
the meson-exchange 
NLSH interaction \cite{MH08}. 
Our interest in this paper is to 
investigate how this phenomenon is modified when the effect beyond the 
mean-field approximation is taken into account.

Fig. 4 shows the projected energy curve, which includes 
the beyond mean-field effect. The solid line shows the energy 
 for the 1/2$^+$ configuration of the $^{31}_{~\Lambda}$Si hypernucleus. 
One can notice that the energy at the spherical configuration is lowered 
when a $\Lambda$ particle is added, as has been indicated also 
in the previous 
mean-field calculations (see also Fig. 3) \cite{MH08,Schulze10}. 
Moreover, the deformation at the energy minimum is shifted towards the 
spherical configuration, that is, 
from $\beta=-0.35$ to $\beta=-0.30$. 
Even though a care must be taken in interpreting the projected energy 
surface, which includes only the rotational correction to the mean-field 
approximation while the vibrational correction is left out \cite{Reinhard78}, 
this may indicate that the collectivity is somewhat reduced in the 
hypernucleus. 

\begin{figure}[t]
\includegraphics[clip,width=8cm]{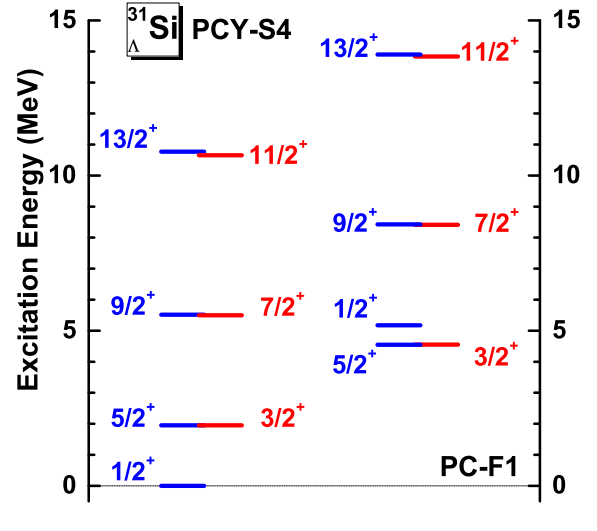}
\caption{
The low-lying spectrum for positive parity states 
of the $^{31}_{~\Lambda}$Si hypernucleus obtained with 
the microscopic particle-rotor model. 
}
\end{figure}

In order to gain a deeper insight into the effect of $\Lambda$ particle 
on the collectivity of the hypernucleus, Fig. 5 shows the spectrum 
of $^{31}_{~\Lambda}$Si for the positive parity states 
obtained with the microscopic particle-rotor model. 
One can observe that the spectrum 
resembles that in the core nucleus shown in Fig. 2. 
These positive parity 
states are in fact dominated by the $\Lambda$ hyperon in the $s$-orbit 
coupled to the positive parity states of the core nucleus. 
However, if one takes the ratio of the energy of the first 4$^+$ state 
to that of the first 2$^+$ state, $R_{4/2}=E(4^+)/E(2^+)$, the addition 
of a $\Lambda$ particle alters it from 
3.083 to 2.829 with the PC-F1 parameter set. 
Here, the ratio for the hypernucleus is estimated 
as $E(9/2_1^+)/E(5/2_1^+)$. 
The $R_{4/2}$ ratio for the core nucleus is close to the value for a rigid 
rotor, that is, $R_{4/2}=3.33$. 
On the other hand, the 
$R_{4/2}$ ratio is significantly reduced in the hypernucleus. 
It is in between the rigid rotor limit and the vibrator limit, that is, 
$R_{4/2}=2.0$, even though the 
$R_{4/2}$ ratio is still somewhat closer to the rigid rotor value. 
This indicates a signature of disappearance of deformation found in the 
previous mean-field calculations \cite{MH08}, even though the deformation 
does not seem to disappear completely and thus the spectrum still shows 
a rotational-like character. 
Of course, the 
weaker polarization effect of a $\Lambda$ particle, which has 
been found also in Ref. \cite{Isaka16}, compared to 
that in the previous mean-field calculations 
is due to the beyond-mean-field effect, that is a combination 
of the effect of shape fluctuation and the angular momentum projection. 
In particular, the 
GCM calculations for the core nucleus indicate that the average 
deformation depends on the angular momentum (see Fig. 1). The impact of 
the $\Lambda$ particle may therefore be state-dependent as well. 

\begin{table}[tb]
\tabcolsep=1.5pt
\caption{The $E2$ transition strengths 
(in units of $e^2$ fm$^4$) for low-lying positive parity states of
$^{30}$Si and $^{31}_{~\Lambda}$Si obtained with the PC-F1 parameter set 
for the $NN$ interaction. 
The c$B(E2)$ values denote 
the corresponding $B(E2)$ values for the core transition in the hypernucleus, 
defined by Eq. (\ref{cBE2}). 
The changes in the $B(E2)$ is indicated with the quantity defined by 
$\Delta\equiv (cB(E2)-B(E2; {^{30}{\rm Si}}))/B(E2; {^{30}{\rm Si}})$.
}
\begin{center}
 \begin{tabular}{cc|ccccc}\hline\hline
 \multicolumn{2}{c|}{$^{30}$Si}& & \multicolumn{4}{c}{$^{31}_{~\Lambda}$Si  }  \\
\hline
$I^\pi_i \to I^\pi_f$  & $B(E2)$   &  &  $J^\pi_i \to J^\pi_f$     
& $B(E2)$ & $cB(E2)$ &    $\Delta$(\%) \\ \hline
 $2^+_1\to 0^+_1$   & 63.60 &  
& $3/2^+_1\to 1/2^+_1$ & 57.00 & 57.00 &  $-$10.38    \\
  & &  & $5/2^+_1\to 1/2^+_1$& 57.06  & 57.06 &  $-$10.28     \\
 $4^+_1\to 2^+_1$    & 103.59  &  & $7/2^+_1\to 3/2^+_1$ 
& 92.14 & 102.38    &  $-$1.17   \\
&  &  & $7/2^+_1\to 5/2^+_1$ & 10.22 & 102.24 &  $-$1.30  \\
&  &  & $9/2^+_1\to 5/2^+_1$ & 102.36 & 102.36 &  $-$1.19  \\
\hline\hline
 \end{tabular}
\end{center}
\end{table}

The calculated quadrupole transition strengths, $B(E2)$, are listed in 
TABLE I. 
Here we also show the $cB(E2)$ values, which are defined as \cite{Mei15}, 
\begin{eqnarray}
cB(E2: I_i\to I_f)&\equiv& 
\frac{1}{(2I_i+1)(2J_f+1)}\,
\left\{
\begin{array}{ccc}
I_f & J_f &  j_\Lambda \\
J_i & I_i &  2 
 \end{array} 
\right\}^{-2} \nonumber \\
&&\times B(E2:J_i\to J_f),
\label{cBE2}
\end{eqnarray}
where $I_i$ and $I_f$ are the dominant angular momenta of the 
core nucleus in the initial and the final hypernuclear configurations, 
while $j_\Lambda$ is that for the $\Lambda$ particle. In the transitions 
shown in TABLE I, $j_\Lambda$ is 1/2. 
This equation is derived by relating 
\begin{eqnarray}
&&B(E2:J_i\to J_f) \nonumber \\
&=&
\frac{1}{2J_i+1}\,
\left|\langle J_i||\hat{T}_{\rm E2}||J_f\rangle\right|^2, \\
&\sim&
\frac{1}{2J_i+1}\,
\left|\left\langle 
[j_\Lambda\otimes I_i]^{(J_i)}
\left|\left|\hat{T}_{\rm E2}\right|\right|
[j_\Lambda\otimes I_f]^{(J_f)}\right\rangle\right|^2, 
\end{eqnarray}
with 
\begin{equation}
B(E2:I_i\to I_f) =
\frac{1}{2I_i+1}\,
\left|\langle I_i||\hat{T}_{\rm E2}||I_f\rangle\right|^2, 
\end{equation}
where $\hat{T}_{\rm E2}$ is the $E2$ transition operator (which acts only 
on the core states). 
The table indicates that the $B(E2)$ transition strengths decrease 
by adding a $\Lambda$ particle into the core nucleus. 
This is consistent with the reduction in deformation in the hypernucleus as 
discussed in the previous paragraph. 

\section{Summary}

We have investigated the role of beyond-mean-field effects on the 
deformation of $^{31}_{~\Lambda}$Si. For this hypernucleus, the previous 
study based on the relativistic mean-field theory had shown that the 
deformation vanishes while the core nucleus, $^{30}$Si, is 
oblately deformed. Using the microscopic particle-rotor model, 
we have shown that the ratio of the energy of the first 4$^+$ state to 
that of the first 2$^+$ state is significantly 
reduced by adding a $\Lambda$ particle to $^{30}$Si, even though the spectrum 
of the hypernucleus 
$^{31}_{~\Lambda}$Si still shows a rotational-like structure. 
This implies that the 
addition of a $\Lambda$ particle to 
$^{30}$Si 
does not lead to a complete disappearance 
of nuclear deformation if the beyond-mean-field effect is taken 
into account, even though the deformation is indeed reduced to some extent. 
In accordance to this, the quardupole transition strengths 
have been found to be also reduced in the hypernucleus. 

Our study in this paper clearly shows that the beyond-mean-field effect 
plays an important role in the structure of hypernuclei. 
We emphasize that 
the microscopic particle-rotor model employed in this paper provides a convenient 
tool for that purpose, which is complementary to the generator coordinate 
method for the whole core+$\Lambda$-particle system \cite{MHY16}. 

\acknowledgments

We thank H. Tamura for useful discussions. 
This work was supported in part by 
JSPS KAKENHI Grant Number 2640263 
and by the National Natural Science Foundation of China under 
Grant No. 11575148.

\end{document}